\newcommand{\rs}{\rho_s}

\newcommand{\ei}{{\eta_i}}

\newcommand{\en}{\epsilon_n}
\newcommand{\ad}{\alpha_d}

\newcommand{\Dp}{{\Delta_\perp}}

\newcommand{\s}{\hat{s}}

\newcommand{\lrb}{L_{\rm RB}}

\newcommand{\pt}{\partial_t}
\newcommand{\Py}{\partial_y}
\newcommand{\ve}{v_E}
\newcommand{\ac}{C_1}
\newcommand{\bc}{C_2}

\documentstyle[twocolumn,aps,graphicx]{revtex}
\begin{document}
\draft
\author{K. Hallatschek, D. Biskamp}
\address{Max-Planck Institut f\"ur Plasmaphysik, EURATOM-IPP
Association, D-85748 Garching, Germany}
\title{Transport control by coherent zonal flows in the
  core/edge transitional regime}
\maketitle
\begin{abstract}
  3D Braginskii turbulence simulations show that the energy flux
  in the core/edge transition region of a tokamak is strongly
  modulated -- locally and on average -- by radially propagating,
  nearly coherent sinusoidal or solitary zonal flows. The flows are
  geodesic acoustic modes (GAM), which are primarily driven by the
  Stringer-Winsor term. The flow amplitude together with the average
  anomalous transport sensitively depend on the GAM frequency and on
  the magnetic curvature acting on the flows, which could be
  influenced in a real tokamak, e.g., by shaping the plasma cross
  section. The local modulation of the turbulence by the flows and the
  excitation of the flows are due to wave-kinetic effects, which have
  been studied for the first time in a turbulence simulation.
\end{abstract}

\paragraph{Introduction ---}
It is now commonly believed, that the transport in the tokamak core is
controlled by zonal flows \cite{diamond98,hahm,terry00}. In the plasma
edge, the flows were not studied thoroughly yet, but they tend to be
weak \cite{zeiler} (although they can possibly completely quench
the turbulence \cite{drakelh}). The zonal flows in the core and at the
edge were found to be incoherent, random fluctuations
\cite{hahm,lin,hallatschek}. In contrast to this, in the transitional
regime between core and edge, strong radially coherent sinusoidal or
solitary zonal flows are ubiquitous according to the numerical
simulations described below. The zonal flows in the transitional
regime are essentially geodesic acoustic modes \cite{gam2}, i.e., the
poloidal rotation is coupled to an $(m,n)=(1,0)$ pressure perturbation
by the inhomogeneous magnetic field, which results in a restoring
force and hence an oscillation. Transcending the present models of the
zonal flow generation based purely on Reynolds stress
\cite{diamond98}, the major part of the flow energy in the
transitional regime is apparently generated by the Stringer-Winsor
term \cite{hassam94}, i.e., the torque on the plasma column caused by
the interaction of pressure inhomogeneities with the inhomogeneous
magnetic field. An order of magnitude variation of the shear flow
level and average transport has been observed upon a sole modification
of the curvature terms acting on the flows, at fixed curvature terms
acting on the turbulence. The pressure inhomogeneities are driven by
anomalous transport modulations, which can be understood by a
drift-wave model for the wave-kinetic turbulence response to the
flows. Some predictions of the model have been verified subsequently
in numerical experiments.

\paragraph*{Zonal flows in the transitional regime are GAMs---}
The numerical turbulence simulations have been carried out using the
three dimensional electrostatic drift Braginskii equations with
isothermal electrons, including the ion temperature fluctuations with
the associated polarization drift effects, the resistive
(non-adiabatic) parallel electron response and the parallel sound
waves (a subset of the equations of Ref.~\cite{drakelh}).
The nondimensional parameters have been varied around a reference
parameter set resembling the transitional core/edge regime: $\ad=0.6$,
$\en=0.08$, $q=3.1$, $\tau=1$, $\eta_i=3$, $\s=1$. The computational
domain is a flux tube winding around the tokamak for three poloidal
connection lengths. The radial and poloidal domain width is $50\lrb$,
with the resistive ballooning scale length, $\lrb$. For a definition
of these parameters and units see Ref.~\cite{drakelh}. The parameters
are consistent with the physical parameters $R=3$~m, $a=1.5$~m,
$L_n=12$~cm, $n=3.5\times10^{19}$~m$^{-3}$, $Z_{\rm eff}=4$,
$B_0=3.5$~T, and $T=200$~eV, $\lrb=3.6$~mm, $\rs=0.82$~mm.  At these
parameters the ITG mode turbulence is the dominant cause of the heat
flux \cite{zeiler}, and the restriction of the domain to a flux
tube is justified \cite{hallatschek}.

\begin{figure}  
(d)\\
\includegraphics[angle=0,width=.8\linewidth]{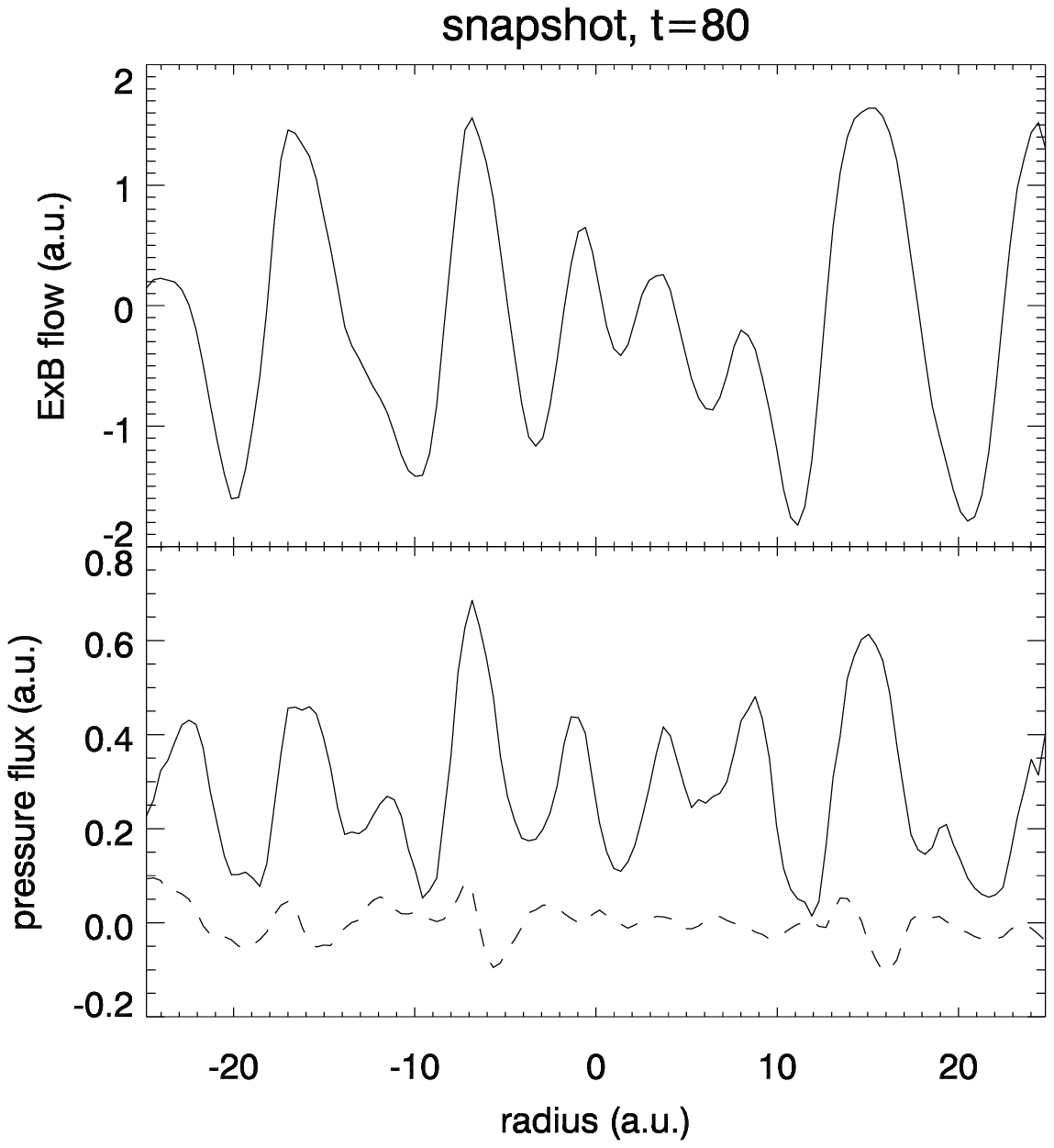}\\ 
\caption{Time evolution of (a) poloidal ${\bf E}\times{\bf B}$ flow
  profile, (b) pressure flux profile $\langle v_r p\rangle$, (c) up-down
  asymmetric pressure flux $\langle v_r p\sin\theta\rangle$ for the
  standard parameters; (d) instantaneous profiles at $t=80$ of
  the flow (above), pressure flux (solid, below), and the up-down
  asymmetric pressure flux (dashed). Note the dipole layers of up-down
  asymmetric pressure flux around the flows, e.g., at $x=-7$. The unipolar
  component of the flux asymmetries can be observed in figure (c):
  depending on their propagation direction the traces look either
  bright or dark.}\label{fig:flow1}
\end{figure}

Viewed as a function of radius and time, the flux surface averaged
poloidal ${\bf E}\times{\bf B}$ flows [fig.~\ref{fig:flow1} (a)] start
as an irregular pattern of radially propagating independent wavelets
and merge into a radially coherent standing wave later on. The final
standing wave pattern consists of GAM oscillations. They can be
described by suitable poloidal Fourier components of the vorticity and
pressure evolution equations (neglecting the parallel sound wave),
\begin{equation}
\partial_t \langle \ve \rangle-\langle s_v \rangle= 
-\ac\langle p \sin \theta\rangle
\label{eq:cba1}
\end{equation}
\begin{equation}
\partial_t\langle p \sin \theta\rangle - \langle s_p \sin \theta\rangle = \bc\frac{\en}{3}
\frac{3+5\tau}{1+\tau}
\langle \ve \sin^2\theta \rangle,\label{eq:cba}
\end{equation}
with $\langle.\rangle$ denoting the flux surface average, the pressure
fluctuations $p=n+\frac{\tau}{1+\tau}T_i$, the poloidal flow velocity
$\ve =\partial_x\phi$, and the source terms $s_v$/$s_p$ of
flow/pressure due to Reynolds stress/anomalous transport,
respectively. For easier reference, the two different curvature terms
in the equations have been adorned with the factors $\ac$ and $\bc$,
which are both $1$ in the turbulence equations for a low aspect ratio
circular tokamak. The curvature term $\ac$ is the Stringer-Winsor
term, the term $\bc$ represents the up-down asymmetric compression of
the plasma due to the poloidal rotation. For all parameters used in
the numerical simulations, the zonal flows oscillate in the stationary
state with a frequency within $5\%$ of the eigenfrequency of
Eqs.~(\ref{eq:cba1},\ref{eq:cba}) without source terms and with the
approximation $\langle \ve \sin^2\theta \rangle\approx\langle \ve
\rangle/2$,
\begin{equation}
\omega=\sqrt{ \ac\bc\frac{\en}{6}\frac{3+5\tau}{1+\tau}}.
\end{equation}
In physical units this is equal to
$\sqrt{(6+10\tau)/(3+3\tau)}c_s/R\sim c_s/R$. Since the
parallel sound frequency $c_s/( q R)$ is much lower than the GAM
frequency, the neglect of the parallel sound wave is justified. The
energy balance equation of the GAM oscillations is according to
(\ref{eq:cba1},\ref{eq:cba})
\[\partial_t\frac{1}{2}\left[\langle \ve \rangle^2
+\frac{\ac}{\omega}\langle p\sin\theta\rangle^2\right]=
\]
\begin{equation}
\langle \ve  \rangle\langle s_v\rangle+\frac{\ac}{\omega}\langle p\sin\theta\rangle\langle s_p\sin\theta\rangle\label{eq:flowenergy}
\end{equation}

The average contributions to the GAM energy from the Reynolds stress
term $\langle s_v\rangle$ and the Stringer-Winsor term $\langle
s_p\sin\theta\rangle$ have been listed in table \ref{tab:contrib} for
varying turbulence parameters. In the transitional regime with its
strong coherent zonal flows most of the flow energy is generated by
the Stringer-Winsor term, while with decreasing temperature toward the
very edge the Stringer-Winsor energy input eventually becomes negative
indicating a braking force on the flows, while simultaneously we get
weak incoherent flows. One is tempted to attribute the decrease of the
zonal flows towards the very edge to the Stringer-Winsor term. Indeed,
eliminating the flow source term $\langle s_p\sin\theta\rangle$ due to
the anomalous transport in the numerical simulations leads to
relatively strong coherent flows even for parameters in the very edge.

With the natural drive of the GAM being apparently the Stringer-Winsor
term, it has been found that altering the amplitude of the curvature
terms $\ac, \bc$ acting on the flows but keeping the curvature terms
acting on the turbulence modes fixed, changes the flow amplitudes and
the anomalous transport by one order of magnitude. Empirically, the
flow level rises with increasing $\ac$ and decreasing $\omega$. One
reason is, that the higher $\ac/\omega\sim\sqrt{\ac/\bc}$ is, the
higher is the contribution of the anomalous transport source term
$s_p$ to the flow energy (\ref{eq:flowenergy}). At constant ratio
$\ac/\omega$ the flows are still somewhat increasing for decreasing
$\omega$. This is understood, since with increasing oscillation period
the flows have more time to influence the turbulence resulting in an
increased $s_p$.

The cause of the pressure perturbations driving the flows are local
modulations of the anomalous transport (the usual Stringer spin-up
mechanism is ineffective due to the relatively long sound transit time
in the considered regime). Plots of the radial pressure transport
$\langle v_r p\rangle=Q(r,t)$ and its up-down antisymmetric component
$\langle v_r p\sin\theta\rangle =U(r,t)$ are shown in
fig.~\ref{fig:flow1} (b) and (c). (Note that the pressure source term
is the divergence of the anomalous pressure flux, i.e.,
$s_p=-\partial_r (v_r p)$, and $\langle
s_p\sin\theta\rangle=-\partial_r U$.) In the initial phase of flow
generation, $U$ develops dipoles around the flows
[fig.~\ref{fig:flow1} (d)], in which $U$ has always the same sign as
the local shearing rate. These dipolar transport structures generate
the pressure up-down asymmetries driving the GAMs. As soon as
sufficiently strong flows exist, the anomalous transport $Q$ develops
a striking peaking at the radii of positive flow resembling transport
fronts propagating with the flow ``waves''. In addition to the dipole
structures $U$ develops a unipolar part component, whose sign depends
on the propagation direction of the corresponding flow. These up-down
antisymmetric transport fronts can be viewed as avalanches running
outward on the lower half of the torus ($\theta<0$) and inward on the
upper half.

The unipolar up-down asymmetries have been found to be responsible for
the setup of the flow pattern. If in a numerical experiment the GAMs
are initially set to zero outside of one flow peak, the turbulence is
still capable of moving this flow into the original direction, until a
new standing wave pattern has been formed. The necessary radial GAM
energy flow due to the turbulence has been found to be primarily
caused by the unipolar up-down asymmetries.  Similar events happen, if
a numerical simulation is started with a GAM pattern with the wrong
wave-length from a simulation run with different GAM parameters. In
that case, strong unipolar up-down transport asymmetries develop,
which enforce the equilibrium flow propagation velocity, i.e., flow
wavelength.

The described peculiar modulation of the transport by the flows is
absent for weak diamagnetic drift and vanishing gyro radius, such as
in the resistive ballooning regime \cite{zeiler}. Instead, the
shear flows simply weaken the turbulence. Apart from this, there is a
tendency to flatten pressure gradients, i.e., to eliminate the
pressure fluctuations associated with the GAM, resulting in the
observed braking force.

\paragraph*{The wave-kinetic effects}
The dependence of the transport modulation on drift effects and the
gyro radius suggests a simplified drift wave model containing only the
radial mode coupling due to the polarization drift, eliminating two
fluid and curvature effects. Since in the numerical studies the mode
wavelengths are not small compared to the zonal flow scales, we
refrain from a geometrical optics approach \cite{diamond98} and
instead move back to the linearized adiabatic drift wave equation,
\begin{equation}
D_t(1-\rs^2\Dp)\phi+\ad\Py \phi=0,\label{eq:drifteq}
\end{equation}
with $D_t=\pt+\ve (x,t)\Py$.  We study the impact of the polarization
drift term $-D_t\rs^2\Dp n$ up to first order in $\rs^2$, without
assumptions on the ratio of flow vs. turbulence scales. The 0-th order
time evolution (with $y$ in Fourier space) is
\begin{equation}
\phi_0(x,t)=\phi_i(x)\psi(x,t)\label{eq:phi0}
\end{equation}
with initial amplitude $\phi_i(x)$ and the flow-and-drift
induced phase factor
\begin{equation}
\psi(x,t)=\exp\left[-ik_y\xi\right], \quad \xi=\int_0^t \left[\ve (x,\tau)+\ad\right] d\tau.\label{eq:xidef}
\end{equation}
Inserting (\ref{eq:phi0}) into (\ref{eq:drifteq}) results in the first
order correction $\phi_1$ due to $\rho_s^2$
\[
\phi_1(x,t)\psi^*(x,t)=-ik_y\ad\rs^2\int_0^t \psi^*(x,\tau)\Dp\left(\phi_i\psi(x,\tau)\right)d\tau+\]
\begin{equation}
+\rs^2 \left[\psi^*(x,\tau)\partial_x^2\left(\phi_i\psi(x,\tau)\right)\right]_{\tau=0}^{\tau=t},
\end{equation}
in which $[f(\dots,\tau)]_{\tau=0}^{\tau=t}\equiv
f(\dots,t)-f(\dots,0)$. From this, the change in turbulence intensity
can be computed to first order,
\[
\delta|\phi|^2=2{\rm Re}(\phi_1^*\phi_0)=2k_y \ad\rho_s^2\int_0^t 
\partial_x[k_x(x,\tau)|\phi_i|^2] d\tau\]
\begin{equation}
-2\rs^2[k_x^2(t)-k_x^2(0)]|\phi_i|^2,
\label{eq:result}
\end{equation}
with a suitable local $k_x(x,\tau)\equiv-k_y\xi'(x,\tau)+{\rm
  Im}(\phi_i'/\phi_i)$. The term involving the diamagnetic drift
velocity $\ad$ corresponds to the advection of mode intensity in
response to the shearing distortion, while the other term is
analogous to the adiabatic compression of a wave field.

\begin{figure}  
\includegraphics[angle=0,width=\linewidth]{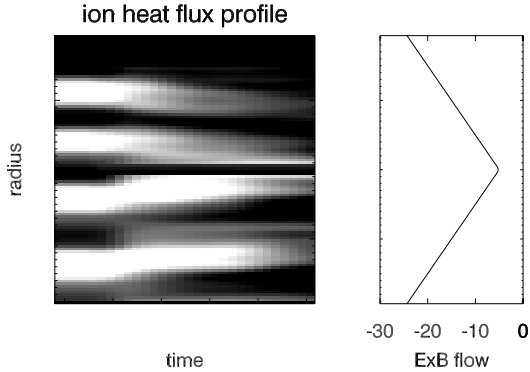}\\ 
\caption{Time evolution of heat flux profile demonstrating
turbulence movement in responce to a linear shear flow with different
sign for the lower and upper half of the plots.}\label{fig:transpadvect}
\end{figure}

If the initial $\phi_i$ has no radial structure ($k_x=0$) and the
shear flows do not change with time, we obtain due to radial mode
advection
\begin{equation}
\delta|\phi|^2_{\rm advection}=-\rs^2\ad|\phi_i|^2t^2k_y^2v''_e,
\end{equation}
which explains the empirical peaking in turbulence intensity at the
locations of positive flows. To verify the presence of the advection
term, a complementary scenario has been simulated, with the turbulence
initially confined to a small region and a constant linear shear flow
$\ve \propto x$ enforced upon it. This results in a motion of the
turbulence maxima towards increasing $\ve $, which confirms the
presence of the advection term (see fig.~\ref{fig:transpadvect}).
In another numerical experiment, a stationary bell shaped flow profile
has been superposed on an initially radially homogeneous turbulence
field, which has been prepared with the zonal flows switched off.
After the turbulence rises transiently at the flow maximum, it drops
to a level below the initial one. This corroborates that the
turbulence amplification at the flow maxima is not due to an increase
of drive at the flow maxima \cite{sidikman94}, but due to a transient
wave-kinetic concentration of fluctuation energy.

The numerically observed radial dipole layers of up-down antisymmetric
transport apparently result from the compressional terms in
(\ref{eq:result}) acting on the mode structure enforced by the
magnetic shear, namely
\begin{equation}
k_x\sim k_y\s\theta.
\end{equation}
Hence, with $\s=1$ a shear flow with positive $v'_e$ reduces $k_x^2$
on the upper side of the tokamak ($\theta>0$) increasing the mode
amplitude there, while the mode is attenuated for ($\theta<0$).
The dependence of the shear flow action on the magnetic shear has been
verified in a numerical experiment with $\s=-1$, in which the modes
are amplified for $\theta v'_e<0$. Moreover, a ``swinging through''
effect hast been observed, i.e., the turbulence intensity decreases
again, when $k_x^2$ rises after $k_x$ has gone through zero.

Last, we consider the unipolar up-down transport asymmetries induced
by a propagating positive shear flow (which is accompanied by a
general transport peaking due to the advective effect). With
$\ve (x,t)=V(x-\nu t)$ (propagation velocity $\nu$) we obtain from
definition (\ref{eq:xidef}) $\xi'(x,t)=-\nu^{-1}V(x-\nu t)$, i.e.
\begin{equation}
k_x(x,t)^2=(k_{x0}-k_y \xi')^2=k_y^2 (\s\theta+\nu^{-1}V(x-\nu t))^2.
\end{equation}
A reduction of $k_x^2$ and corresponding amplification of the
turbulence modes via eq.~(\ref{eq:result}) occurs for $\s\theta\nu
V<0$. The sign of this effect of a moving flow has been confirmed
numerically for positive and negative shear.

\paragraph*{Conclusions ---}
GAMs, oscillating zonal flows, have been found to be the main
mechanism controlling the turbulence level in the transitional
core/edge regime. Quasi-stationary zonal flows are unimportant in the
simulations, because of the strong restoring force due to the pressure
imbalance generated by the magnetic field inhomogeneity, which cannot
be short-circuited along the magnetic field lines due to the long
parallel sound transit time in the transitional regime. Primarily, the
flows are not driven by Reynolds stress but by the pressure
asymmetries on a flux surface generated by modulations of the
anomalous transport. These modulations in turn are caused by the
flows, on one hand by the radial advection of turbulence energy
concentrating the turbulence in locations of increased flow in
electron diamagnetic direction, and on the other hand by adiabatic
compression effects on the wave field which together with the magnetic
shear result in up-down asymmetries of the transport. The peculiar
nature of the drive mechanism leads to propagating peaked flow
structures acompanied by pronounced transport fronts, which sometimes
come close to propagating solitons. We have studied for the first time
the action of wave-kinetic effects in a numerical simulation.  Similar
results should be expected for the core, except that the polarization
due to the ion larmor radius has to be replaced by the neoclassical
polarization due to the banana width \cite{rosenbluth}.  I.e., we
expect the wave-kinetic transport modulation in the core to be
significantly stronger than in in the edge. GAMs on the other hand are
less important in the core due to the shorter parallel connection
length ($q\sim1$).

On one hand, the GAM drive efficency depends on the nature of the
turbulence, in that it depends on the presence of finite $\rs$
effects. Absence of these, such as in the resistive ballooning regime
leads to a strong damping of the GAM due to the anomalous diffusion
eroding the pressure fluctuations connected with the GAM.  However,
apart from the turbulence the zonal flow amplitude is influenced
strongly by the linear properties of the GAM itself, such as its
frequency or the torque excerted on a pressure fluctuation.
Manipulation of both of them can lead to a reduction of the transport
by up to an order of magnitude. Consequentially, any discussion of the
anomalous transport focusing on the influence of the magnetic geometry
on the turbulence drive falls short of an essential factor, if the
influence of the magnetic geometry on the flows is neglected.  Since
the GAM properties could be influenced in a real tokamak by, e.g.,
shaping the plasma column, the coherent flows should be taken into
consideration to further reduce the transport in advanced tokamaks.
Moreover, the clear signature of the radially coherent flows in the
numerical simulations makes them an interesting target of experimental
investigation, e.g., by means of microwave reflectometry.

This work has been performed under the auspices of the
Center for Interdisciplinary Plasma Science, a joint initiative by the
Max-Planck-Institutes for Plasma Physics and for Extraterrestrial Physics.

\begin{figure}  
(a)\\
\includegraphics[angle=0,width=.8\linewidth]{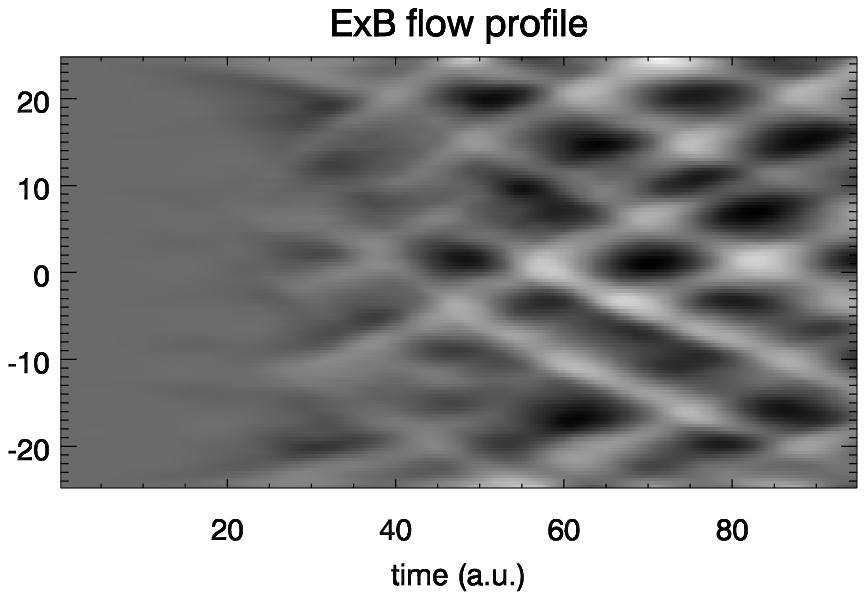}\\ 
\end{figure}
\begin{figure}  
(b)\\
\includegraphics[angle=0,width=.8\linewidth]{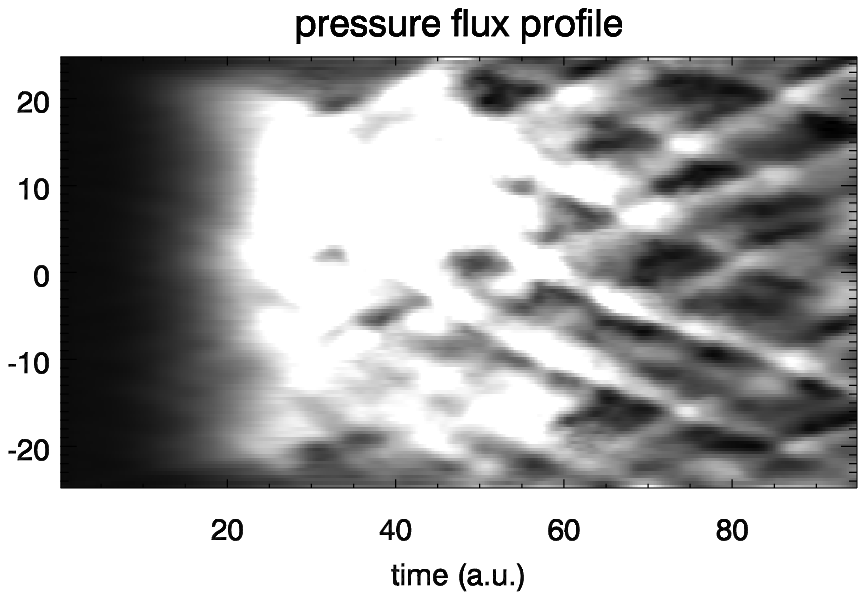}\\ 
\end{figure}
\begin{figure}  
(c)\\
\includegraphics[angle=0,width=.8\linewidth]{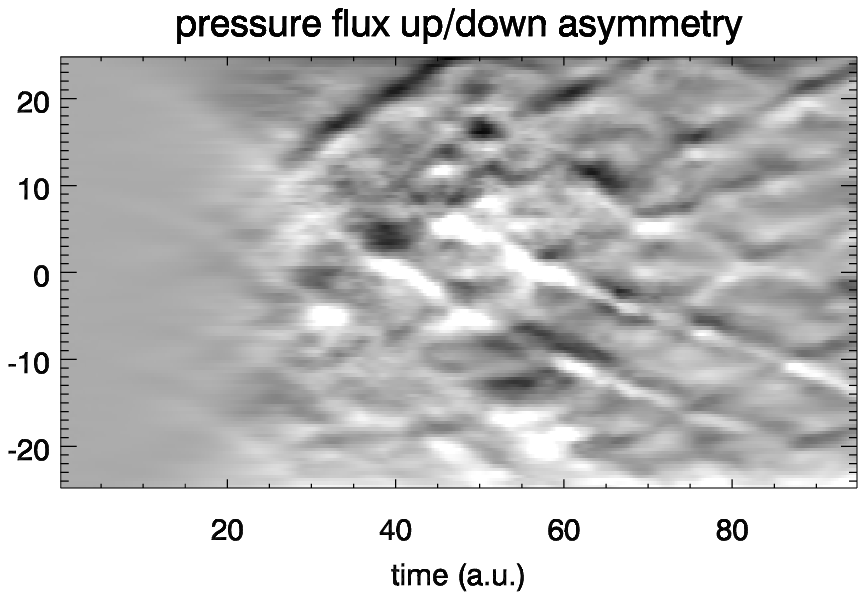}\\ 
\end{figure}
%
%
\begin{table}
  \caption{Mean GAM energy production due to Reynolds stress and
    Stringer-Winsor effect, flow intensity $\langle \ve ^2\rangle$
    and anomalous pressure transport $\langle p
    v_r\rangle$ for varying parameters. The units are the turbulence
    units described in reference {\protect\cite{drakelh}}. In these units, the
    diamagnetic velocity is equal to $\ad$. The
    parameters are consistent with the physical reference parameters
    given in the text, except for an altered temperature
    $T_i=T_e=T$. For $\ad=0$ the ions have been assumed cold. In the
    line marked with an $*$ the Stringer-Winsor drive has been
    switched off, eliminating its braking force in the ballooning regime.}
\label{tab:contrib}
\begin{tabular}{ccccccc}
$\ad$&$\ei$&$T$ (eV)&$\langle p v_r\rangle$&$\langle \ve ^2\rangle$&
Reynolds drive&Stringer drive\\\hline
0    &$-$&  0&0.27&0.04&$1.4$&$-1.0$\\
0$^*$&$-$&  0&0.25&0.2  &3.3&$-26$\\
0.2  &1&100&0.27&0.03& 2.0 &$-1.0$\\
0.2  &3&100&0.87&0.23& 1.5 &3.6\\
0.6  &3&300&0.15&0.9&$-0.1$ &10.0\\ 
1.1  &3&550&0.24&0.9 &16.0 &33.4\\
\hline
\end{tabular}
\end{table}
\end{document}